# Bitwise Logic Using Phase Change Memory Devices Based on the Pinatubo Architecture

Noa Aflalo, Eilam Yalon, and Shahar Kvatinsky
The Andrew and Erna Viterbi Faculty of Electrical and Computer Engineering
Technion – Israel Institute of Technology
Haifa, Israel 3200003
noa.aflalo@campus.technion.ac.il, eilamy@technion.ac.il, shahar@ee.technion.ac.il

*Abstract*—This paper experimentally demonstrates a near-crossbar memory logic technique called Pinatubo. Pinatubo, an acronym for Processing In Non-volatile memory ArchiTecture for bUlk Bitwise Operations, facilitates the concurrent activation of two or more rows, enabling bitwise operations such as OR, AND, XOR, and NOT on the activated rows. We implement Pinatubo using phase change memory (PCM) and compare our experimental results with the simulated data from the original Pinatubo study. Our findings highlight a significant four-orders of magnitude difference between resistance states, suggesting the robustness of the Pinatubo architecture with PCM technology.

## I. Introduction

Throughout the past century, computers have traditionally separated the memory and processing unit. This architecture, known as von Neumann architecture, was initially introduced in 1945 by John von Neumann. As the technology evolved, processing units saw remarkable advancements in speed and power. Yet, the ability to transfer data from the memory to the processing unit lagged behind. This disparity is referred to as the "memory wall" [2-5].

One strategy to address the memory wall issue is through the adoption of *Processing-in-Memory* (PIM), shown in Fig. 1, which facilitates computation within the memory unit by incorporating digital logic into it [6-8]. PIM offers numerous benefits, including high bandwidth, extreme parallelism, and impressive energy efficiency. Emerging nonvolatile memristive technologies serve dual roles, acting as both the memory and logical units to facilitate PIM [9-10].

Stateful logic and non-stateful logic are two methodologies for leveraging memristive devices in PIM architectures, each integrating the processing unit with data storage. In stateful logic, data processing and storage share the same representation domain (e.g., resistance is used in memristors for data storage and as the inputs and outputs of the gate). In non-stateful logic, the computation is performed near or within the memory array, while the logic state is represented as voltage or current [11-14].

This paper examines Pinatubo [1], an architecture facilitating near-memory non-stateful logic. The Pinatubo architecture supports various emerging memristive technologies and significantly enhances data-intensive graph processing and bitwise operations compared to traditional processors.

Phase Change Memory (PCM) represents one form of memristive technology. Constructed from phase-change materials, PCM can hold data in both amorphous and crystalline states. A PCM device operates with a unipolar switching mechanism, and phase transitions are triggered by the application of an external voltage. This voltage induces a thermal shift in the phase change material, causing it to alternate between amorphous and crystalline states [15-18].

The cross-sectional schematic of a confined PCM is shown in Fig. 2a. It has a bottom electrode, then an insulator with a programmable region in the middle. On top, it has the phase change material and the top electrode to apply voltage. PCM devices support three operations: read, set, and reset, as illustrated in Fig. 2b. A read operation is conducted at a relatively low bias to avoid self-heating and maintain the phase of the PCM device. A set operation is performed with a medium amplitude pulse that heats the material to an intermediate temperature, higher than the crystallization temperature (~150 C). The set pulse should be sufficiently long (~ 0.1-10 μs), causing the phase change material to set into the crystalline phase. A reset operation is executed with a fast (<50 ns) and high-amplitude pulse-inducing temperature higher than the melt (~600 C), which resets the phase change material into the amorphous phase [18-19].

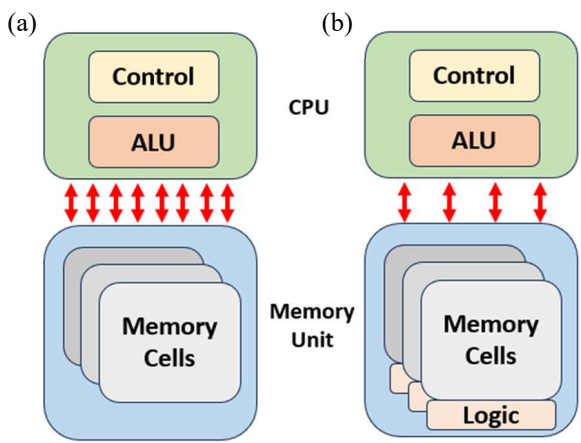

Fig. 1. (a) von Neumann architecture. Computing-centric architecture as the main memory, seperated from the CPU. (b) PIM architecture with processing unit in the memory which eliminates data transfer to the CPU by perfomning logical operations within the memory.

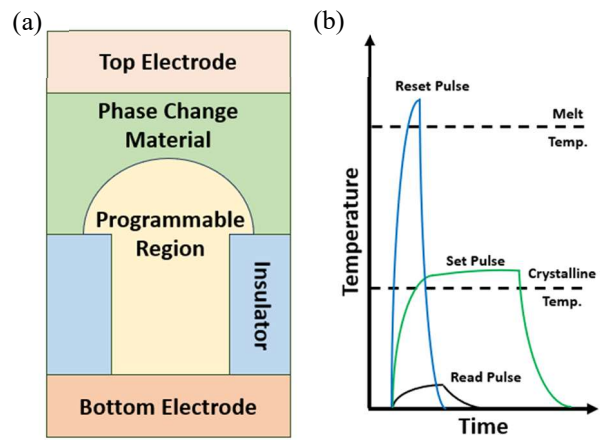

Fig. 2. Phase Change Memory. (a) Cross-sectional schematic of the conventional PCM. (b) Read, set, and reset operations. single step. Crystallization temperature is ~150 C, melt temperature ~600C.

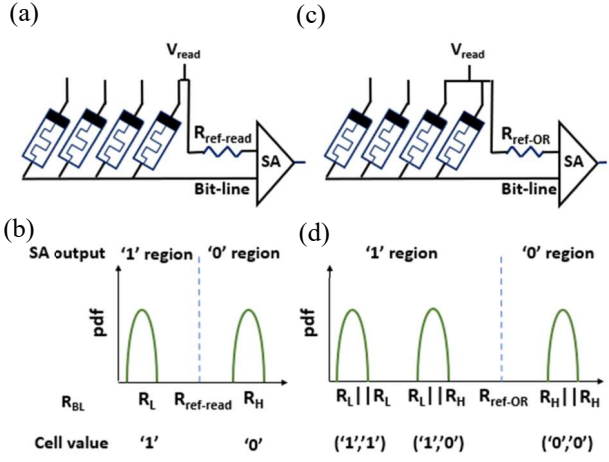

Fig. 3. Pinatubo schematic and modification of reference resistance values for the SA of OR operation. (a) Schematic for SA read using $R_{ref-read}$. (b) SA read using $R_{ref-read}$ different regions. (c) Schematic for SA computes an OR using $R_{ref-OR}$. (d) SA computes an OR using $R_{ref-OR}$ different regions.

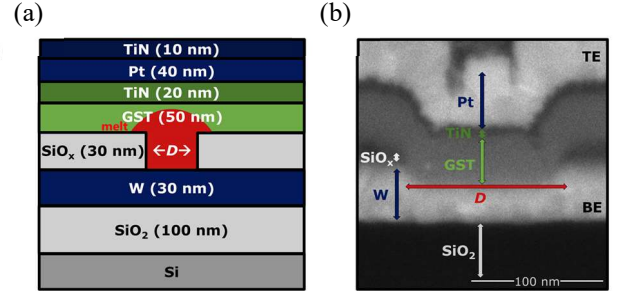

Fig. 4. PCM devices used in this work. (a) Cross section schematic. (b) Cross section scanning electron microscopy (SEM). The tungsten (W) layer was patterned by evaporation and etching. GST layer, the TiN/Pt TE, and contact pads was patterned by sputtering and liftoff. D~125 nm [25].

TABLE I. OR BITWISE OPERATION

| OR Bitwise Operation | | | | | | |
|---|---|---|---|---|---|---|
| Logical State | | | Resistance State | | | OR |
| One cell | Two cells | Three cells | One cell | Two cells | Three cells | |
| ('1') | ('1','1') ('1','0') | ('1','1','1') ('1','1','0') ('1','0','0') | ($R_L$) | ($R_L,R_L$) ($R_L,R_H$) | ($R_L,R_L,R_L$) ($R_L,R_L,R_H$) ($R_L,R_H,R_H$) | '1' |
| ('0') | ('0','0') | ('0','0','0') | ($R_H$) | ($R_H,R_H$) | ($R_H,R_H,R_H$) | '0' |

This paper provides experimental validation of the near-memory logic technique of Pinatubo using PCM. We demonstrate accurate logic behavior for two and three-input logic gates. The high OFF-ON ratio in PCM devices ensures reliable operation.

## II. PINATUBO

Pinatubo is an architecture proposed to accelerate bitwise operations using nonvolatile memories (NVMs). Bitwise operations are very common and used widely in databases, graph processing, bioinformatics, and image processing. They can be used to replace complex arithmetic operations [1], [21-24].

Pinatubo is a PIM architecture, and the core components of the design are shown in Fig. 1. Fig. 1a shows a common approach to computing data stored in memory. Each bit-vector is fetched sequentially from the memory to the computing unit, namely, the processor, then the data is executed in the ALU, and lastly, the data is written back to the memory. Fig. 1b shows the PIM approach, where the computation is done in the memory unit. Only commands and addresses are sent from the processor to memory. The bitwise operations in Pinatubo are performed by activating two (or more) memory rows and passing the results through a sense amplifier (SA). Then, the results are written to the destination address in the memory unit. Pinatubo allows the reduction of data movement, high internal bandwidth, and massive parallelism all at once [1].

The key idea of Pinatubo is using a SA for bitwise operations, as shown in Fig. 3. The resistance of a single cell ($R_L$ or $R_H$) is compared with a reference value ($R_{ref-read}$), as shown in Fig. 3a. Fig. 3b shows the difference between cell storing logical '0' and a cell storing logical '1' regions using comparing with $R_{ref-read}$. When activating two cells, the resistance in the bit-line connected to the SA is the parallel resistance of these two cells, as shown in Fig. 3c. In this case, there will be three resistances to differentiate between: $R_L||R_L$, $R_L||R_H$, and $R_H||R_H$, and Fig. 3d shows the differentiation between the resistances. For implementing OR operation, the reference resistance ($R_{ref-OR}$) will be in between $R_L \| R_H$ to $R_H \| R_H$. The same method can be modified for more cells. To compute other bitwise operations, the reference resistance needs to be modified. Table 1 lists the notations for OR bitwise operation in terms of logical state and resistance state for one/two/three cells [1].

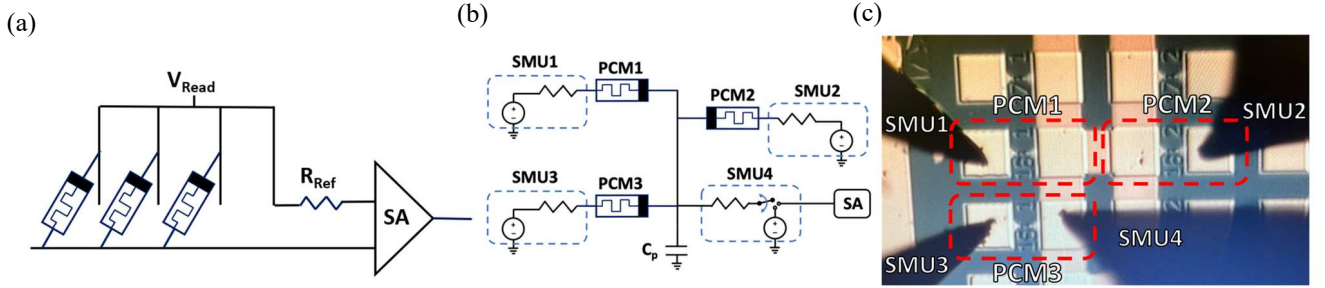

Fig. 5. Experimental setup. (a) Pinatubo basic operation. (b). Schematic top view measurement setup. (c) Optical top view of the devices. Three devices are connected in parallel with their bottom electrode shared with the bit-line, and their top electrode is connected to WGFMU channel to force the read voltage. The bit-line is connected to the fourth WGFMU channel to sense the current.

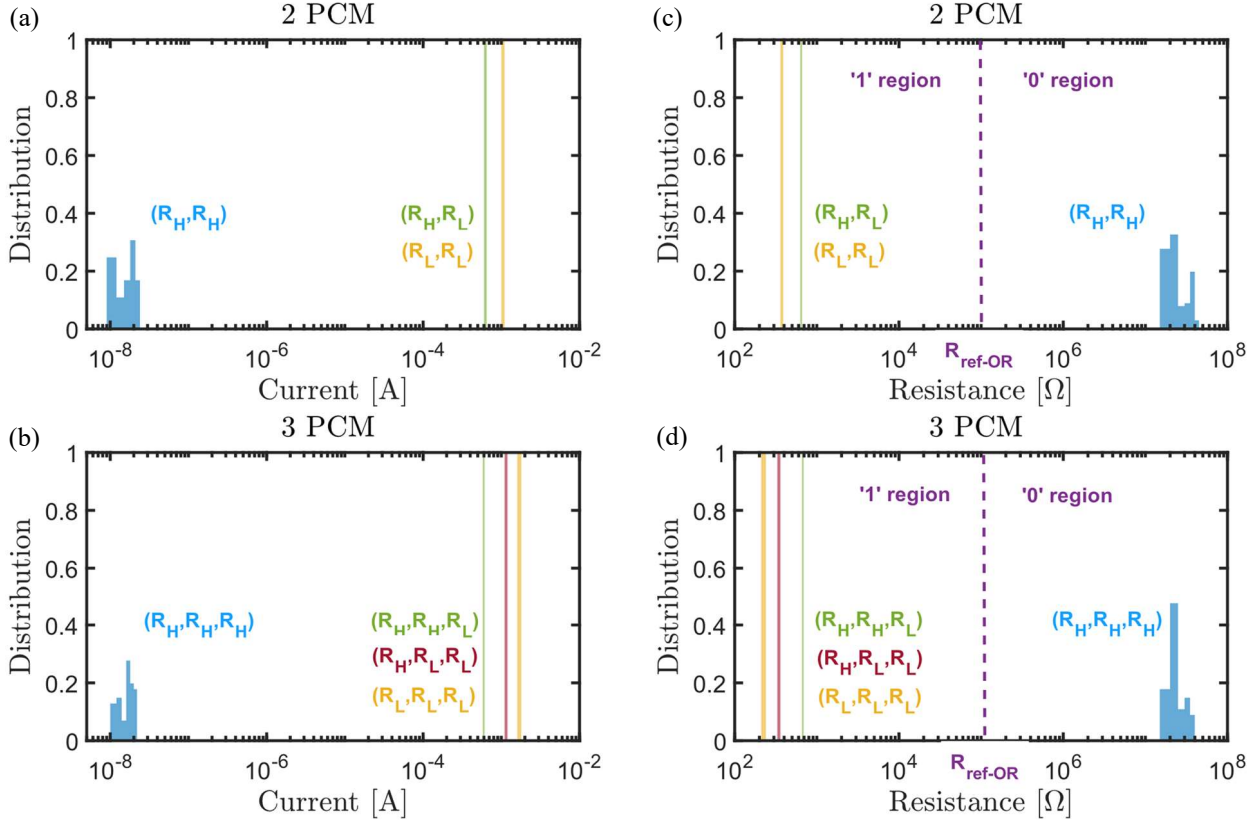

Fig. 6. Experimental results. (a) Current measurements of two PCM. (b) Current measurements of three PCM. Each state was measured 100 times. (c) Pinatubo regions and $R_{ref-OR}$ for the OR gate for 2 PCM devices and (d) 3 PCM devices.

### III. EXPERIMENTS

#### A. Device Fabrication

The PCM devices used in this work are similar devices to those used in [25]. The device structure and the materials being used are shown in Fig. 4. The tungsten (W) layer was patterned by evaporation and etching. GST layer, the TiN/Pt TE, and contact pads was patterned by sputtering and liftoff.

#### B. Electrical Measurements

The electrical measurements were performed using Keysight B1500A with four WGFMU channels and two probe channels. The measurement setup has three PCM devices, a sense amplifier, and a voltage source. The output of the bitwise operation is the output of the SA as illustrated in Fig. 5a. The three PCM devices are connected in parallel to the bit-line, then the bit-line is connected to the SA, and the other port of the SA is connected to the reference resistance. Fig. 5b and Fig. 5c show the devices that were measured in a schematic and optical top view. Three devices are connected in parallel with their bottom electrode to the bit-line, and their top electrode is connected to WGFMU channel to force the read voltage. The bit-line is connected to the fourth WGFMU channel to sense the current.

#### C. Read, Set, and Reset Operations

Two WGFMU channels are employed to execute set and reset operations, as depicted in Fig. 2b, facilitating the transition between amorphous and crystalline phases. The set pulse was designed with an amplitude of 2.7 V, a rise time of 10 ns, a duration of 30 ns, and a fall time of 3 μs. The reset pulse featured an amplitude of 4 V, with a rise time, duration, and fall time of 10 ns each. For read operations, two standard probe channels were utilized. An applied voltage of 0.4 V was used, and the resulting current was measured.

#### D. Experimental Logic Demonstration

We have experimentally validated the Pinatubo logic for two and three PCM devices connected in parallel. Each experiment was performed 100 times, using sequences of set, reset, and read operations as detailed below. The measurements were performed repeatedly on the same devices. An extended reliability study, including a higher number of cycles and an analysis of device to device variability will be part of future work.

The configuration illustrated in Fig. 5b was employed for both experimental setups. The two-device experiments involved PCM1 and PCM2, assigned to channels 1 and 2, respectively. For the three-device experiment, an additional device, PCM3, was incorporated with an extra channel. Initially, all devices were reset to a high resistive state using the reset operation. Subsequently, a voltage of 0.4 V was applied using channels 1 and 2 (including channel 3 for the three-device experiment), and the resultant current was measured via channel 4. After measuring the high resistive state of both PCM devices (both inputs are logical 0), both devices were set to the low resistive state. The same reading mechanism was used. For measuring other input combinations, the inputs in logical 0 were reset and the inputs in logical 1 were set.

The results of both experiments are shown in Fig. 6. For both experiments, using two and three parallel devices, the measured current when all inputs are in the low resistive state is approximately 1 mA, and the measured current when all inputs are in the high resistive state is approximately 10 nA.

The results from both experiments correlate with the resistance distributions expected when using Pinatubo architecture, as shown in Fig. 3. The distinct regions are easily distinguishable, and the selection of $R_{ref\text{-}OR}$ can be readily determined, as indicated in Fig. 6c and Fig. 6d. A reference resistance of $R_{ref\text{-}OR}=100$ kΩ proves adequate for both 2-PCM and 3-PCM operations. This approach demonstrates robustness, given the four orders of magnitude difference between the different input combinations. Hence, we deduce from these experiments that by selecting an appropriate reference current, we can differentiate various logical states and perform the OR operation.

Conversely, selecting $R_{ref\text{-}AND}$ for an AND bitwise operation poses a greater challenge due to the less than one order of magnitude difference between the different input combinations. This minor difference complicates the process of choosing a proper resistance value for the reference resistor, particularly when accounting for variations.

We do not include here a quantified analysis of the energy consumption, because such analysis should be conducted at the system level, whereas our measurements were performed at the device level.

IV. CONCLUSION

This paper presented an implementation for Pinatubo architecture, employing PCM devices as nonvolatile memory. We demonstrated an OR bitwise operation utilizing PCM devices and a sense amplifier. A remarkable four orders of magnitude difference between various input combinations were reported, underscoring the robustness of the Pinatubo architecture when implemented with PCM devices. This architecture holds significant potential in PIM architectures to address the speed and power limitations imposed by the memory wall.

ACKNOWLEDGMENT

Fabrication was performed at the Stanford Nanofabrication Facility (SNF), Stanford, CA, USA, and the Technion Micro-Nano Fabrication Unit (MNFU). This project has received funding from the European Union's Horizon 2020 Research And Innovation Programme FET-Open NEU-Chip under grant agreement No. 964877. The authors thank Orian Leitersdorf for his feedback throughout this work.